\documentclass[journal]{IEEEtran}
\usepackage{cite}     
\usepackage{graphicx}

\usepackage{amsmath}

\begin{document}
\title{Scintillators with Silicon Photomultiplier Readout for Timing Measurements in Hadronic Showers}

\author
{
{

Christian Soldner for the CALICE Collaboration

}
}

\maketitle
\thispagestyle{empty}

\begin{abstract}
The advent of silicon photomultipliers has enabled big advances in high energy
physics instrumentation, for example by allowing the construction of extremely
granular hadronic calorimeters with photon sensors integrated into small
scintillator tiles. Direct coupling of the SiPM to the plastic scintillator,
without use of wavelength shifting fibers, provides a fast detector response,
making such devices well suited for precise timing measurements. We have
constructed a setup consisting of 15 such scintillator tiles read out with fast
digitizers with deep buffers to measure the time structure of signals in
hadronic calorimeters. Specialized data reconstruction algorithms that allow the
determination of the arrival time of individual photons by a detailed analysis
of the recorded waveforms and that provide automatic calibration of the gain of
the photon sensor, have been developed. We will discuss the experimental
apparatus and the data analysis. In addition, we will report on first results
obtained in a hadronic calorimeter with tungsten absorbers, providing important
constraints on the time development of hadronic showers for the development of
simulation models.
\end{abstract}

\begin{IEEEkeywords}
Silicon Photomultiplier, Scintillator, Calorimetry, Timing
\end{IEEEkeywords}

\maketitle
\thispagestyle{empty}

\section{Introduction}

Multi-cell Geiger-mode avalanche photodiodes (G-APDs), often referred to as
Silicon Photomultipiers (SiPMs)\cite{Bondarenko:2000in} have a wide range of
applications in high energy physics instrumentation. They provide high photon
detection efficiency and insensitivity to magnetic fields with very compact
devices. A large number of such devices, approximately 8\,000, have been
successfully used to read out small plastic scintillator tiles in the CALICE
analog hadron calorimeter\cite{Adloff:2010hb}, a physics prototype of a highly
granular calorimeter for detectors at the future International Linear Collider
(ILC). In addition to the proof of technology, the test beam campaigns with this
and other calorimeter prototypes also yield information on the structure of
hadronic showers with unprecedented precision, providing important input for the
further development of hadronic shower models used in simulation tools for high
energy physics and beyond.

The development of detector concepts for the Compact Linear Collider (CLIC)
\cite{Assmann:CLIC} has led to
considerable interest in using tungsten instead of steel absorbers also in the
hadronic calorimeters, since this allows compact detectors. At CLIC, time stamping of
signals will be of key importance, requiring the study of the intrinsic time
structure of hadronic showers in a tungsten calorimeter with scintillator
readout, to evaluate the impact of physical processes on the achievable time
resolution.

\section{Experimental Setup}

\begin{figure}
\centering
\includegraphics[width=0.47\textwidth]{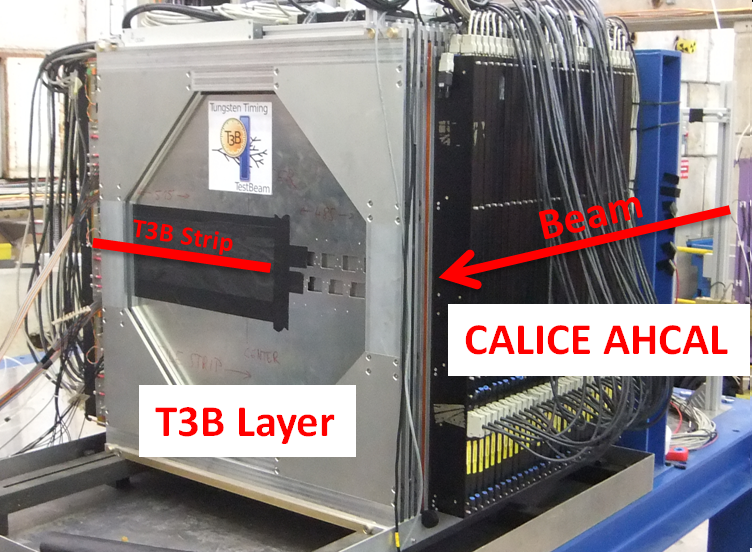}
\caption{The CALICE analog tungsten HCAL at the CERN PS, with the T3B setup installed in the last layer of the detector.}
\label{fig:T3BInstallation}
\end{figure}

\begin{figure}
\centering
\includegraphics[width=0.49\textwidth]{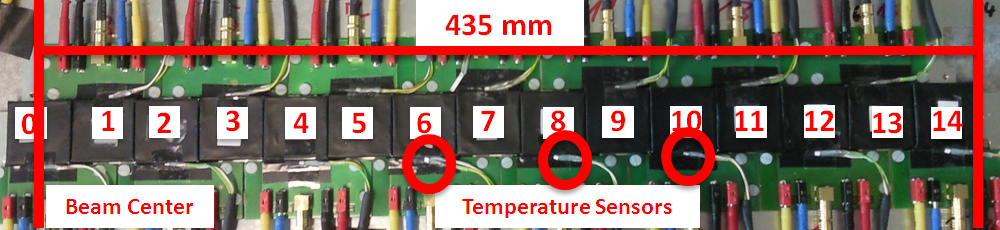}
\caption{Layout of the 15 T3B scintillator tiles, covering the radial shower
profile from the beam line to the outer region. Each tile is equipped with a
sensor which monitors temperature variations close to the SiPM coupling
position.}
\label{fig:T3BTileLayout}
\end{figure}

To provide first measurements of the time structure of hadronic showers in a
tungsten calorimeter with scintillator readout, a dedicated apparatus, the T3B
(Tungsten Timing Test Beam) experiment, has been constructed. This device is
operated together with the main CALICE analog scintillator tungsten HCAL, as an
additional active layer behind the main detector. The setup consists of 15
scintillator cells with a size of $3\,\times\,3$ cm$^2$ and a thickness of 5 mm,
with directly coupled photon sensors. As direct coupling requires the
sensitivity maximum of the sensors to match the light emission maximum of
conventional scintillators,
blue sensitive SiPMs developed by
Hamamatsu\footnote{Hamamatsu photonics (http://www.hamamatsu.com/)} were used.
Without wavelength-shifting fibers, these cells provide a fast response to ionizing
particles. A uniform response across the whole surface area is provided by a
special shaping of the coupling position, as discussed in \cite{Simon:2010hf}.
The 15 T3B cells are arranged in one row extending from the center of the
calorimeter out to one side of the detector (Figure \ref{fig:T3BInstallation})
and positioned in a depth of 4 nuclear interaction lengths. The analog SiPM
signals are read out with four 4-channel
USB-oscilloscopes\footnote{PicoTech PicoScope 6403 (http://www.picotech.com/)}
which provide a sampling rate of 1.25\,GSa/s on all channels and are therefore
well suited for precise timing measurements in the nanosecond region. Long
acquisition windows of 2.4$\mu s$ per event are recorded to study the time
structure of the energy deposits in the scintillator in detail, providing
information on the time structure of hadronic showers in the calorimeter.
 
Shower events are triggered synchronous with the CALICE HCAL to allow for an
event correlation of T3B to CALICE data. In the main calorimeter, the position
of the first inelastic hadronic interaction can be determined event by event,
allowing to measure the time structure of the shower at various depths with
respect to the shower start, which can be used to measure the averaged timing
profile over the full longitudinal and lateral extent of the shower.

The T3B detector was part of the CALICE test beam campaign at the CERN Proton
Synchrotron in 2010 and at the Super Proton Synchrotron in 2011, and
successfully acquired large data sets of hadronic showers in an energy range of
2-300\,GeV.

\section{Data Reconstruction}

\begin{figure}
\centering
\includegraphics[width=0.49\textwidth]{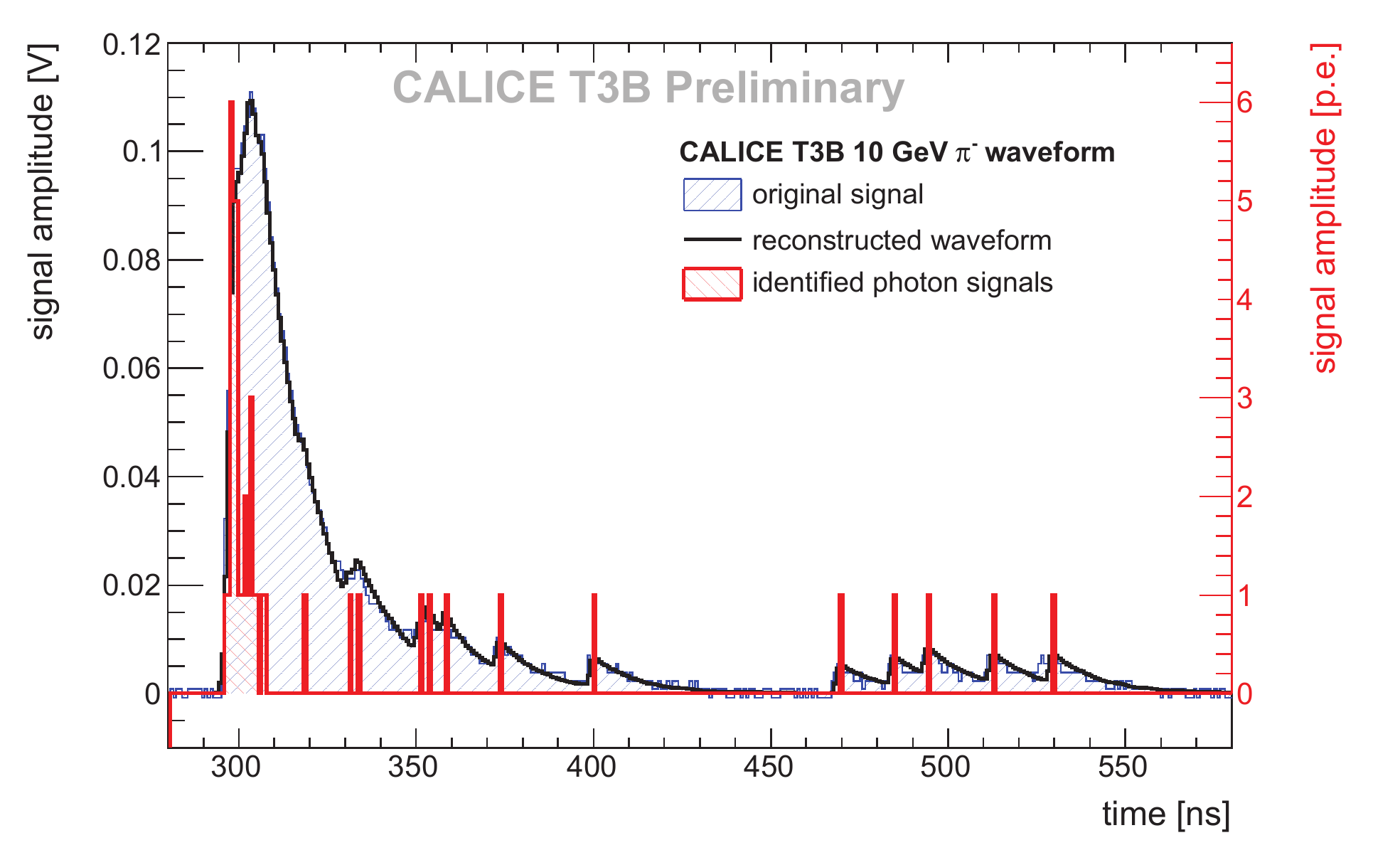}
\caption{Typical waveform of a T3B shower event on one channel. The original
SiPM signal (blue) is iteratively decomposed into its single photon
contributions using the averaged 1 photon signal from calibration data. The
obtained time distribution of single photon hits (red) can be used to
reconstruct the original waveform (black) and validate the success of the
procedure.}
\label{fig:RecoWaveForm}
\end{figure}

The intrinsic properties of SiPMs require a dedicated set of calibration tools.
A specialized data reconstruction algorithm was developed to determine the
arrival time of individual photons at the sensor with sub-nanosecond precision
eliminating any influence of the temperature dependence of the SiPM gain at
the same time.

At test beam facilities, a certain number of particles is delivered to the
physics experiments within a time window of a few seconds (for the SPS)
in so-called spills, leaving the experiments time for readout and calibration
before the next particle spill arrives. The T3B detector recorded physics events
during the spill and SiPM dark count events in between spills
for live calibration purposes of the corresponding physics events.

As a first step in the calibration sequence, zero suppression based on pedestals
determined on a spill-by-spill basis was applied to the aquired SiPM signals.

Then, the signals were decomposed into individual photon equivalents. Selecting
single pixel dark count waveforms from the inter-spill calibration data and
averaging the waveforms on a cell-by-cell basis over 10 spills corresponding to
less than 5 minutes, one obtains a reference signal for a single SiPM geiger
discharge or a single detected photon respectively. This averaged calibration
waveform is then iteratively subtracted from local maxima detected in the
corresponding physics waveforms until no maxima above 0.5 p.e. remain. Figure
\ref{fig:RecoWaveForm} shows one example of a waveform decomposed using this
reconstruction technique. To check the quality of this analysis, a waveform
based on the identified photon signals was built up with the reference single
photon signals and compared to the original waveform. The very good agreement
between measurement and the reconstructed waveform demonstrates the quality of
reconstruction.

The calibration procedure results in an implicit gain calibration, since
possible cell-to-cell gain differences lead to corresponding differences in the
average single photon signals used in the calibration. The resulting number of
photons is thus independent of the SiPM gain. Due to the continuous automatic
updating of the average single photon waveforms, the T3B detector is not
affected by SiPM gain variations due to temperature changes and the temperature
dependence of the measured signal amplitudes is significantly reduced.
Additionally, the calibration procedure results in a data reduction factor of
approximately 1000. Any further data analysis was executed on waveform
decomposed T3B events using the timing of each identified photon.

\begin{figure}
\centering
\includegraphics[width=0.49\textwidth]{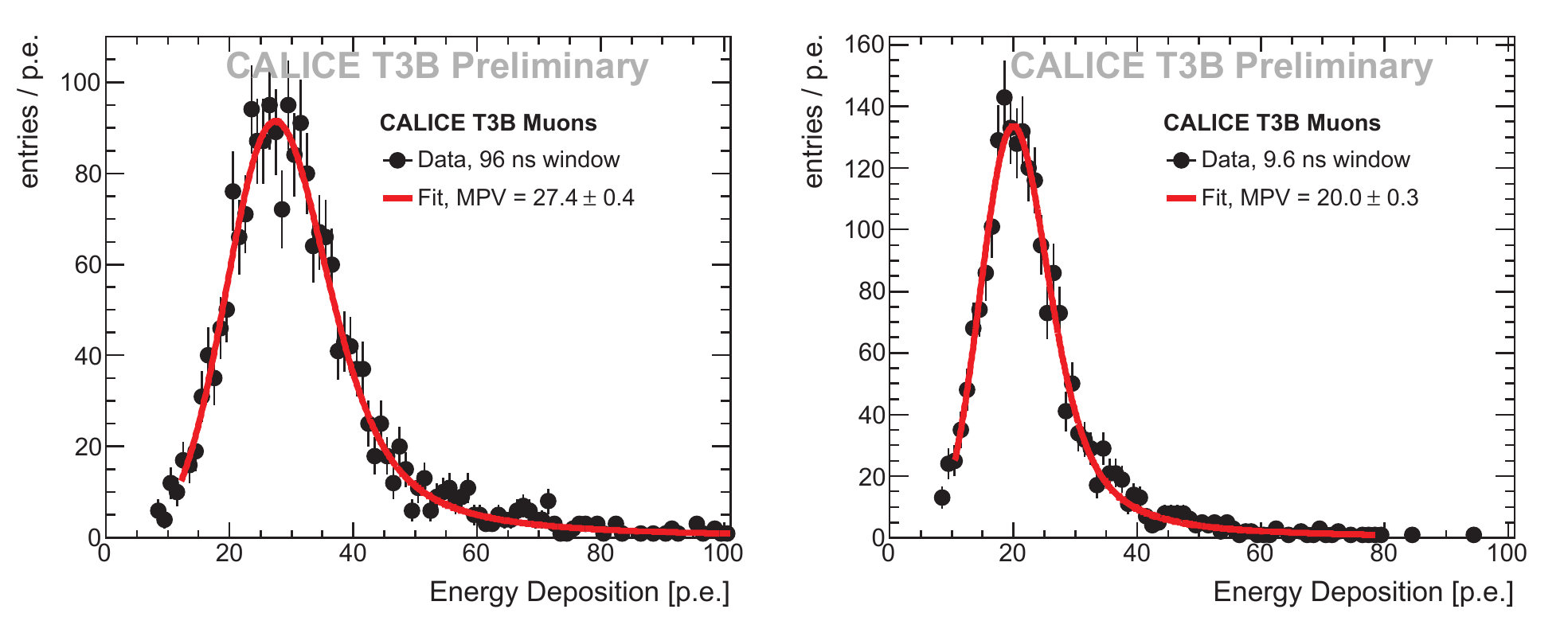}
\caption{Measured spectrum for muons in the central T3B scintillator tile,
reconstructed by identifying the time of individual photon signals in the SiPM,
for two different integration time windows: 96 ns from the first identified
photon (left) and a time window of 9.6 ns (right). The distributions were fitted
with a Landau function convolved with a Gaussian to extract the most probable
value. In both cases, the $\chi^2$ per degree of freedom is around 0.8,
indicating a good fit quality.}
\label{fig:Muons}
\end{figure}

Figure \ref{fig:Muons} shows the distribution of the energy reconstructed with
this technique in the central tile of the T3B detector for muons obtained in a
special data run with an absorber in the beam line. The deposited energy was determined with
two different integration windows, 96 ns from the first identified photons,
shown in Figure \ref{fig:Muons} {\it left} and 9.6 ns, shown in Figure
\ref{fig:Muons} {\it right}. The most probable value of both distributions was
extracted by fitting a Landau function convoluted with a Gaussian. The
integration time has a considerable effect on the most probable value, which is
reduced by almost 30\% from $27.4\,\pm\,0.4$ p.e. to $20.0\,\pm\,0.3$ p.e. for
the reduced integration window. Also the width of the signal is reduced
considerably. This is due to the partial exclusion of contributions from
afterpulses of the photon sensor.
 
Further calibration algorithms are to be applied to address thermal darkrate and
signal afterpulsing and to calibrate energy depositions to the scale of minimum
ionizing particles.

\begin{figure}
\centering
\includegraphics[width=0.42\textwidth]{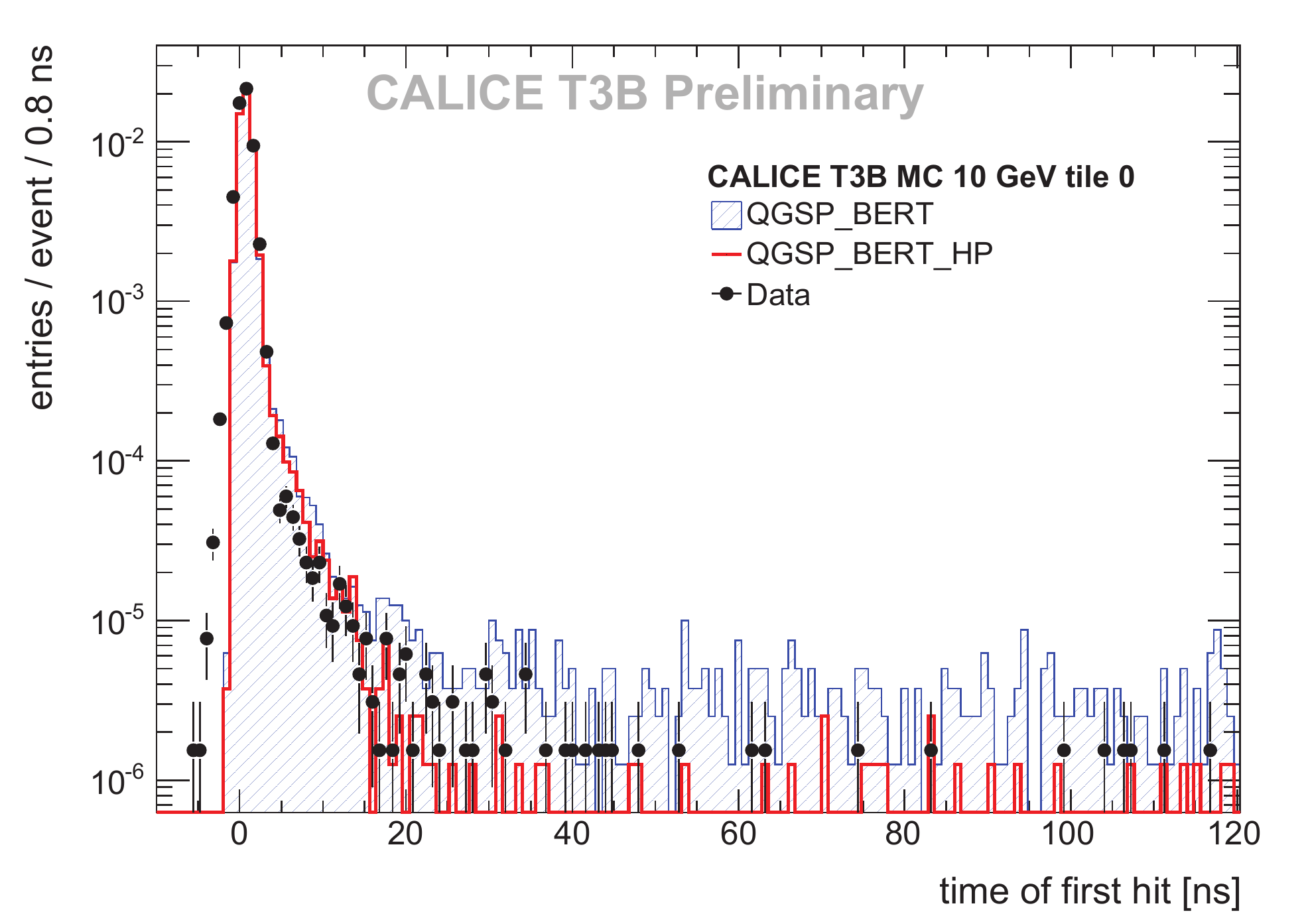}
\caption{Comparison of the time of first hit distribution in the central T3B
tile simulated for 10\,GeV $\pi^-$ with QGSP\_BERT (blue) and QGSP\_BERT\_HP
(red) with data (black).}
\label{fig:TimeDistribution}
\end{figure}

\section{First Results}
As a first parameter in the investigation of the time structure of hadronic
showers, the time of the first energy deposit is studied. This parameter
provides a good indication of the intrinsic time stamping possibilities in the
calorimeter. The analysis is performed on the decomposed waveforms for a 10\,GeV
$\pi^-$ run with 718\,000 events. A SiPM signal is considered as energy
deposition where at least 8 photon equivalents are detected within 12 time bins
(9.6 ns). The time of hit is then taken from the timing of the second detected
photon of that deposition. It was observed that using the first instead of the
second photon leads to additional jitter due to single p.e. dark counts before
the starting time of the real hit. The distribution of the time of first hit for
the central T3B tile is shown in Figure \ref{fig:TimeDistribution}. A simulation
study, based on a direct implementation of the CALICE tungsten HCAL geometry
together with an approximation of the T3B detector in GEANT4
\cite{Agostinelli:GEANT4} was performed to provide first comparisons to the T3B
results. The data was compared to two different hadronic shower models,
QGSP\_BERT, and QGSP\_BERT\_HP \cite{Ribbon:QGSP_BERT_HP}, a variant which
provides additional high precision neutron tracking, and is expected to give an
improved description of the shower evolution in heavy absorbers, while the
former is the physics list mostly used in the simulation of LHC detectors and
for linear collider optimization studies. Figure \ref{fig:TimeDistribution}
illustrates a striking difference in the late shower evolution for the two
GEANT4 physics models. The delayed energy deposits are considerably reduced in
the model including high precision neutron tracking and this model is consistent
with the observation in data. This is most prominent in the tail of the hit
distribution beyond 20 ns. The main peak is reasonably well described by both
models. The standard QGSP\_BERT list on the other hand overestimates isolated
late energy depositions, where isolated means that only the time of the first
hit is taken into account in case of multiple hits on one tile in the same
event.

\begin{figure}
\centering
\includegraphics[width=0.42\textwidth]{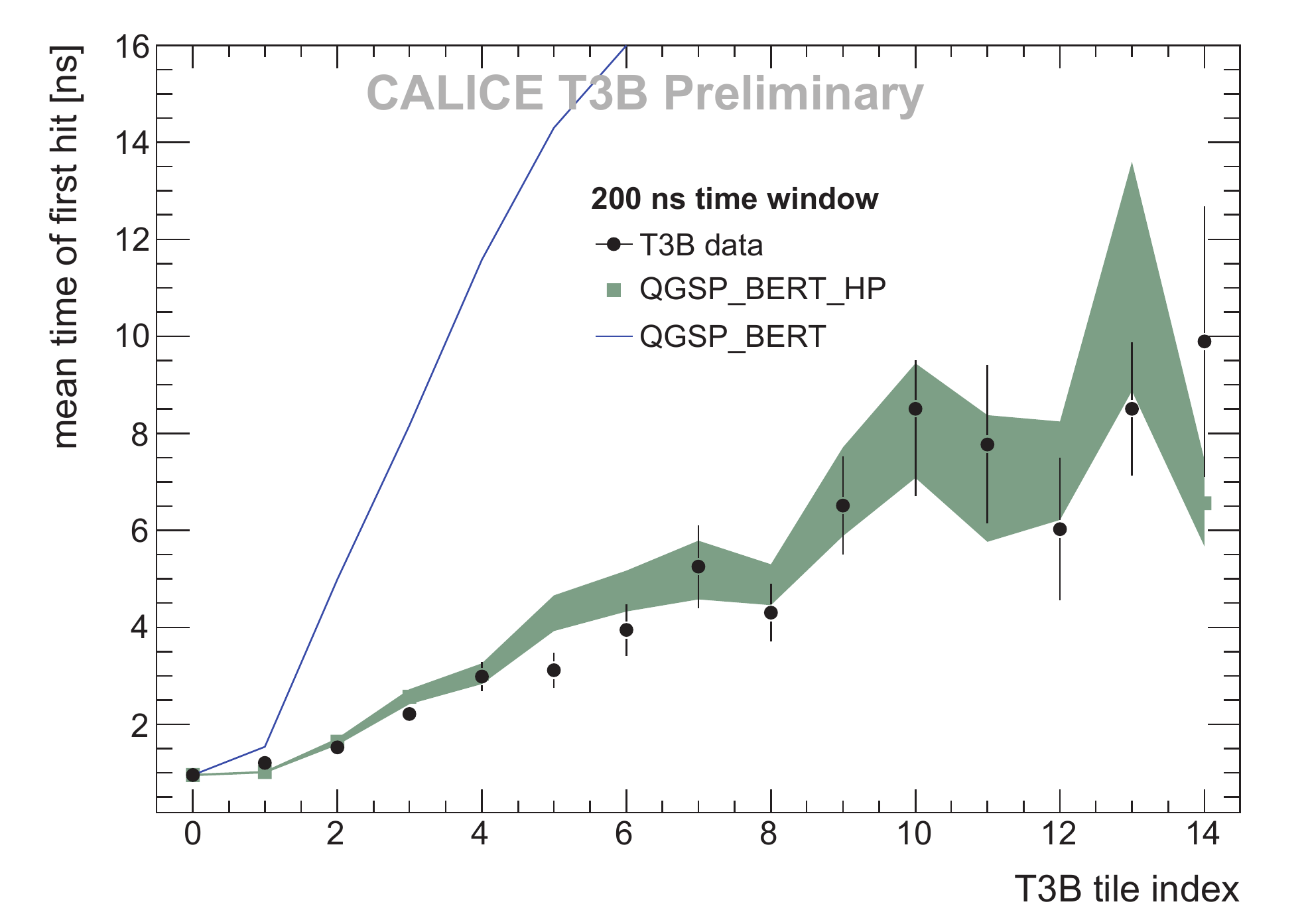}
\caption{Mean time of first hit for 10 GeV $\pi^-$ as a function of radial
distance from the shower core (a tile index of 10 corresponds to approximately
30 cm). The data are compared with simulations using \texttt{QGSP\_BERT} and
\texttt{QGSP\_BERT\_HP}. The error bars and the width of the area in the case of
\texttt{QGSP\_BERT\_HP} simulations show the statistical error, while for 
\texttt{QGSP\_BERT} the errors are omitted for clarity.}
\label{fig:MeanTofH}
\end{figure}

To provide first information on the lateral shape of the time development of
hadronic showers, the mean time of first hit for each of the T3B
cells was determined from the distribution of the first hit discussed above. The
mean was formed within a time window of 200 ns, starting 10 ns before the
maximum of the distribution in T3B tile 0. This time window covers the time
relevant for calorimetry at CLIC, where the duration for one bunch train is
expected to be 156 ns. Figure \ref{fig:MeanTofH} shows the mean time of first
hit as a function of the radial distance from the shower axis. The beam axis
passes through T3B tile 0, so that a tile index of 10 corresponds to a distance
of approximately 30 cm. While \texttt{QGSP\_BERT\_HP} gives an excellent
description of the data, \texttt{QGSP\_BERT} shows very large discrepancies,
with significantly overestimated late contributions at larger radii.

This demonstrates the importance of the high precision neutron tracking in
GEANT4 for a realistic reproduction of the time evolution of hadronic showers in
tungsten. Upcoming T3B results will investigate the time development of hadronic
showers in detail and provide more information that can be used for the
validation and further development of the timing aspect of shower models in
GEANT4.

\bibliographystyle{IEEEtran.bst}
\bibliography{CALICE}

\end{document}